\documentclass[aps,twocolumn,superscriptaddress]{revtex4-1}
\usepackage[colorlinks=true,bookmarks=true,citecolor=magenta,urlcolor=magenta,linkcolor=magenta,breaklinks]{hyperref} 
\usepackage{breakurl}
\usepackage{sidecap}
\usepackage{amssymb}
\usepackage{hhline}
\usepackage{urwchancal}
\usepackage{xcolor}
\sidecaptionvpos{figure}{t}
\usepackage{amsmath}
\usepackage{graphicx}
\usepackage{esint}
\usepackage{epstopdf}
\usepackage{rotating}
\graphicspath{{pict/}{./}}
\usepackage{bm}%
\usepackage{microtype,bm,bbm,graphicx,booktabs,times}

\newcounter{Fig}

\begin{document}


\title{Scattering invariance for arbitrary polarizations protected by joint spatial-duality symmetries}
\author{Qingdong Yang}
\affiliation{School of Optical and Electronic Information, Huazhong University of Science and Technology, Wuhan, Hubei 430074, P. R. China}
\author{Weijin Chen}
\affiliation{School of Optical and Electronic Information, Huazhong University of Science and Technology, Wuhan, Hubei 430074, P. R. China}
\author{Yuntian Chen}
\email{yuntian@hust.edu.cn}
\affiliation{School of Optical and Electronic Information, Huazhong University of Science and Technology, Wuhan, Hubei 430074, P. R. China}
\affiliation{Wuhan National Laboratory for Optoelectronics, Huazhong University of Science and Technology, Wuhan, Hubei 430074, P. R. China}
\author{Wei Liu}
\email{wei.liu.pku@gmail.com}
\affiliation{College for Advanced Interdisciplinary Studies, National University of Defense
Technology, Changsha, Hunan 410073, P. R. China}

\begin{abstract}
We reveal how to exploit joint spatial-electromagnetic duality symmetries to obtain invariant scattering properties (including extinction, scattering, absorption) of self-dual scattering systems for incident waves of arbitrary polarizations. The electromagnetic duality ensures the helicity preservation along all scattering directions, and thus intrinsically eliminates  the interferences between the two scattering channels originating from the circularly polarized components of incident waves. This absence of interference directly secures invariant scattering properties for all polarizations located on the same latitude circle of the Poincar\'{e} sphere, which are characterized by polarization ellipses of the same eccentricity and handedness. Further incorporations of mirror and/or inversion symmetries would lead to such invariance throughout the whole Poincar\'{e} sphere,  guaranteeing invariant scattering properties for all polarizations. Simultaneous exploitations of composite symmetries of different natures render an extra dimension of freedom for scattering manipulations, offering new insights for both fundamental explorations and optical device engineering related to symmetry dictated light-matter interactions.
\end{abstract}

\maketitle

\section{Introduction}
\label{section1}
Like in many other disciplines of physics, the principles of symmetries have been pervasive and widely exploited in different branches of photonics~\cite{FEYNMAN_2011__Feynmanb,LANDAU_1984__Electrodynamicsb,JACKSON_1998__Classical,Joannopoulos2008_book,BARRON_2009__Molecular}. For various optical elements, the optical responses are to a large extent decided by the scattering features of their irreducible constituents, flexible manipulations of which are crucial for not only fundamental studies but also practical optical device designs~\cite{Bohren1983_book,LIU_2005__Photonic}. This makes the related branch of Mie scattering essential and ubiquitous throughout photonics, where a tight grasp of the scattering properties of building atoms serves widely as the cornerstone for further explorations.  For many specific applications, stable optical functionalities that are immune to inevitable polarization fluctuations in realistic devices are desired. This can be largely reduced to another classical problem in Mie theory: \textit{how to obtain inherently invariant scattering properties for arbitrary polarizations?}

It has been revealed that generic scattering invariance can be obtained based on pure spatial symmetries of scattering configurations, not only for some specific sets of polarizations (\textit{e.g.} linear polarizations~\cite{Hopkins2013_nanoscale} or circular polarizations~\cite{CHEN_2020_Phys.Rev.Research_Scatteringa}), but also for all polarizations covering the whole Poincar\'{e} sphere that is widely employed for arbitrary polarization characterization~\cite{YANG_2020_ArXiv200613466Phys._Symmetry}. Besides spatial symmetries, it is recently shown that alternatively the electromagnetic duality symmetry~\cite{JACKSON_1998__Classical,FERNANDEZ-CORBATON_2013_Phys.Rev.Lett._Electromagnetica} can be also exploited to render similar scattering invariance for linear polarizations~\cite{MOHAMMADIESTAKHRI_2020_Phys.Rev.Lett._Electromagnetic,YANG_2020_ACSPhotonics_Electromagnetic,YANG_2020_ArXiv200610629Phys._Scattering}. Compared to spatial symmetries that can be easily broken when incident waves are also taken into consideration~\cite{CHEN_2020_Phys.Rev.Research_Scatteringa}, duality symmetry is intrinsic (decided by the scattering bodies only and not affected by the external incident waves) and thus generally intact for all incident directions, despite that it imposes more stringent restrictions on the optical properties of the scattering systems~\cite{FERNANDEZ-CORBATON_2013_Phys.Rev.Lett._Electromagnetica,FERNANDEZ-CORBATON_2013_Phys.Rev.B_Role,RAHIMZADEGAN_2018_Phys.Rev.Applied_CoreShella,MOHAMMADIESTAKHRI_2020_Phys.Rev.Lett._Electromagnetic,YANG_2020_ACSPhotonics_Electromagnetic,YANG_2020_ArXiv200610629Phys._Scattering}.
It has been demonstrated that combined spatial-duality symmetry can be employed for manipulations of angular scattering patterns~\cite{FERNANDEZ-CORBATON_2013_Opt.ExpressOE_Forwarda,ZAMBRANA-PUYALTO2013Opt.Lett.,MOHAMMADIESTAKHRI_2020_Phys.Rev.Lett._Electromagnetic,YANG_2020_ACSPhotonics_Electromagnetic}, and is thus natural to expect that such joint symmetries can be further exploited to obtain all-polarization independent scattering properties.

Here we unveil how discrete spatial symmetries and electromagnetic duality symmetry can be exploited simultaneously to guarantee invariant scattering properties (including cross sections of extinction, scattering and absorption) for arbitrary incident polarizations. It is revealed that for arbitrary self-dual scattering bodies lacking any spatial symmetry, sole duality symmetry is sufficient to ensure that the two scattering channels (contributed by circularly-polarized components with opposite handedness of the incident waves) are effectively decoupled and orthogonal to each other along all directions.  This leads to scattering invariance for polarizations on the same latitude circle of the Poincar\'{e} sphere, that is, polarizations of the same ellipse eccentricity and handedness. Further incorporation of mirror and/or inversion symmetries can further extinguish the optical activities (distinct responses for circular polarizations of opposite handedness), result in invariant scattering properties for arbitrary polarizations. The symmetry principles we have revealed can bring new perspectives for manipulations of symmetry-dictated light matter interactions, benefiting both symmetry-based fundamental explorations (\textit{e.g.} topological photonics and parity-time symmetry optics) and optical device engineering.

\section{Helicity Preservation along arbitrary scattering directions for self-dual scattering systems}
\label{section2}

Self-dual scattering systems are those that are invariant under duality transformations~\cite{JACKSON_1998__Classical,FERNANDEZ-CORBATON_2013_Phys.Rev.Lett._Electromagnetica,YANG_2020_ACSPhotonics_Electromagnetic,YANG_2020_ArXiv200610629Phys._Scattering}, which in essence cannot differentiate the electric and magnetic components of incident electromagnetic waves.  It has been proved that for self-dual scattering bodies, along all scattering directions there is helicity preservation (including those directions where there are no radiations, for which the helicities are not well-defined)~\cite{FERNANDEZ-CORBATON_2013_Phys.Rev.Lett._Electromagnetica,FERNANDEZ-CORBATON_2013_Opt.ExpressOE_Forwarda,CHEN_ACSOmega_Global}.  It means that, for incident left- and right-handed circularly polarized (LCP and RCP) waves, the scattering along any direction is also respectively LCP and RCP. In this section, we aim to provide a much simpler, more intuitive and accessible proof to verify this all-angle helicity preservation feature, which will serve as a foundation for our further discussions of scattering invariance for arbitrary polarizations.

As a first step, we assume that the incident plane wave is propagating along \textbf{z} direction [wave vector \textbf{k} $\|$ \textbf{z}; see also the coordinate system shown in Fig.~\ref{fig1}(a)], which can be expanded into either linear basis \textbf{X} and \textbf{Y} (linearly polarized wave along \textbf{x} and \textbf{y} axis, respectively) or circular basis \textbf{L} and \textbf{R} (LCP and RCP waves, respectively).  When the incident wave is \textbf{X}, we denote the scattered wave along an arbitrary direction \textbf{s} in circular basis as:
\begin{equation}
\label{scattering-arbitrary}
\mathbf{E}_s=\alpha (\mathbf{s}) \mathbf{{L}}+ \beta (\mathbf{s}) \mathbf{{R}},
\end{equation}
where $\alpha (\mathbf{s})$ and $\beta (\mathbf{s})$ are complex expansion coefficients.  Then we implement a $\pi/2$ duality transformation for the whole scattering configuration (self-dual scattering body is invariant upon duality transformations; incident wave and scattered waves are rotated by $\pi/2$ along \textbf{k} and \textbf{s}, respectively)~\cite{YANG_2020_ACSPhotonics_Electromagnetic}, which would convert the incident wave from \textbf{X} to \textbf{Y},  and the scattered wave in Eq.~(\ref{scattering-arbitrary}) to~\cite{YANG_2020_ArXiv200613466Phys._Symmetry}:
\begin{equation}
\label{scattering-arbitrary-transform}
\mathbf{E}_s=-i\alpha (\mathbf{s}) \mathbf{{L}}+ i\beta (\mathbf{s}) \mathbf{{R}}.
\end{equation}
where the extra phase term $\pm i$ for RCP and LCP scattered components originates from the nontrivial spin rotation phase of circularly-polarized light, which has been discussed in detail in Feynman Lectures (Volume III, Chapter 11)~\cite{FEYNMAN_2011__Feynmanb3}. For incident waves \textbf{L} and \textbf{R}, or equivalently $1/\sqrt{2}(\mathbf{X} \pm i\mathbf{Y})$, according to Eqs.~(\ref{scattering-arbitrary}) and (\ref{scattering-arbitrary-transform}) the scattered waves are:
\begin{equation}
\label{circular-scattering}
\mathbf{E}_{s}=\frac{1}{\sqrt{2}}[(\alpha (\mathbf{s}) (1 \mp i^2)\mathbf{L}+\beta(\mathbf{s}) (1 \pm i^2) \mathbf{R}],
\end{equation}
which are exactly LCP and RCP, respectively ($1+i^2=0$). This concludes the proof that the helicity is preserved along arbitrary scattering directions for self-dual scattering bodies. This helicity preservation feature is much stronger than the rotation symmetry induced one discussed in Ref.~\cite{FERNANDEZ-CORBATON_2013_Opt.ExpressOE_Forwarda,YANG_2020_ArXiv200613466Phys._Symmetry}, where there is only helicity preservation along the forward scattering direction.

\section{Invariant scattering properties for arbitrary polarizations protected by joint spatial-duality symmetries}
\label{section3}

\subsection{General Theoretical Analysis Based on Duality Symmetry Principles}
\label{section3-1}
Similar to Eq.~(\ref{scattering-arbitrary}),  incident waves of arbitrary polarizations can be expressed as:
\begin{equation}
\label{incident-arbitrary}
\mathbf{E}_i=\alpha_i \mathbf{{L}}+ \beta_i \mathbf{{R}},
\end{equation}
and the scattered waves  are related to the incident waves through the linear scattering matrix $\hat{\mathbf{T}}$~\cite{Bohren1983_book,YANG_2020_ArXiv200613466Phys._Symmetry}:
\begin{equation}
\label{matrix}
\mathbf{E}_{s}=\hat{\mathbf{T}}\mathbf{E}_{i}=\alpha_i \mathbf{E}_{s}^{L}+\beta_i \mathbf{E}_{s}^{R},
\end{equation}
where $ \mathbf{E}_{s}^{L}=\hat{\mathbf{T}}\mathbf{{L}}$ and $ \mathbf{E}_{s}^{R}=\hat{\mathbf{T}}\mathbf{{R}}$ are scattered waves with incident waves \textbf{L} and \textbf{R}, respectively. According to the helicity preservation discussed in Section~\ref{section2}, $\mathbf{E}_{s}^{{L}}$ and $\mathbf{E}_{s}^{{R}}$ are intrinsically LCP and RCP waves along all directions, which are orthogonal to each other and thus there is no interference between those two circularly polarized scattering channels.  As a result, the angular scattering intensity can be directly expressed as:
\begin{equation}
\label{angular-intensity}
\rm{I}_{s}=|\alpha_i|^2 \rm{I}_{s}^{L}+|\beta_i|^2 \rm{I}_{s}^{R},
\end{equation}
where $\rm{I}_{s}^{L}=|\mathbf{E}_{s}^{{L}}|^2$ and $\rm{I}_{s}^{R}=|\mathbf{E}_{s}^{{R}}|^2$ are angular scattering intensities with incident waves \textbf{L} and \textbf{R}, respectively.

The optical theorem~\cite{Bohren1983_book} tells that the extinction is decided by interferences between incident wave $\mathbf{E}_i$ and forward scattered wave $\mathbf{E}_s (\mathbf{s}\| \mathbf{k})$.  Since there is no cross interference between LCP and RCP components, the helicity preservation along the forward direction directly guarantees that~\cite{YANG_2020_ArXiv200613466Phys._Symmetry}:
\begin{equation}
\label{extinction}
\rm C_{\rm{ext}}=|\alpha_i|^2 \rm C_{\rm{ext}}^L+|\beta_i|^2 \rm C_{\rm{ext}}^R,
\end{equation}
where  $\rm {C}_{\rm{ext}}^L$ and $\rm {C}_{\rm{ext}}^R$ are extinction cross sections for incident waves \textbf{L} and \textbf{R}, respectively. The scattering cross section $\rm C_{\rm{sca}}$ can be obtained by integrating the angular scattering intensity shown in Eq.~(\ref{angular-intensity}). Since there is no interference along any direction for the two scattering channels $\mathbf{E}_{s}^{{L}}$ and $\mathbf{E}_{s}^{{R}}$, the scattering cross section can be in a similar fashion expressed as:
\begin{equation}
\label{scattering-cross}
\rm C_{\rm{sca}}=|\alpha_i|^2 C_{\rm{sca}}^L+|\beta_i|^2 \rm C_{\rm{sca}}^R,
\end{equation}
where  $\rm {C}_{\rm{sca}}^L$ and $\rm {C}_{\rm{sca}}^R$ are scattering cross sections for incident waves \textbf{L} and \textbf{R}, respectively.  Also according to the optical theorem~\cite{Bohren1983_book}, the absorption cross section can be obtained by subtracting scattering cross section from the extinction cross section $\rm C_{\rm{abs}}=\rm C_{\rm{ext}}-\rm C_{\rm{sca}}$, leading to the following expression absorption cross section:
\begin{equation}
\label{absorption}
\rm C_{\rm{abs}}=|\alpha_i|^2 \rm C_{\rm{abs}}^L+|\beta_i|^2 \rm C_{\rm{abs}}^R,
\end{equation}
where  $\rm {C}_{\rm{abs}}^L$ and $\rm {C}_{\rm{abs}}^R$ are absorption cross sections for incident waves \textbf{L} and \textbf{R}, respectively.

For convenience of discussions, we define the circular component ratio for the arbitrarily-polarized incident wave [refer to Eq.~(\ref{incident-arbitrary})] as $\gamma_i=\beta_i/\alpha_i$: its amplitude $|\gamma_i|$ decides the eccentricity and handedness of the polarization ellipse, while its phase Arg($\gamma_i$) decides the orientation of the polarization ellipse~\cite{YANG_2020_ArXiv200613466Phys._Symmetry,BERRYM.V._2003_Proceeding_optical}.  A constant $|\gamma_i|$ corresponds to the polarizations on the same latitude circle of the Poincar\'{e} sphere, that is, polarizations characterized by ellipses of the same eccentricity and handedness~\cite{YANG_2020_ArXiv200613466Phys._Symmetry}. Two singular positions without well defined phase are $|\gamma_i|=0,~\infty$, which corresponds respectively to LCP and RCP waves (ellipse orientations are not defined). Linear polarizations constitute a special case of this with $|\gamma_i|=1$, which locate on the equator of the Poincar\'{e} sphere.

According to Eqs.~(\ref{angular-intensity})-(\ref{absorption}), the scattering properties of self-dual scattering bodies have nothing to do with  Arg($\gamma_i$), and thus are fully independent of the orientations of the incident polarization ellipses (\textit{e.g.} the polarization directions of the incident linearly polarized light).   Those equations also tell convincingly that to achieve invariant scattering properties (in terms of both cross sections and angular scattering patterns), at least one of the following conditions has to be satisfied: (i) $|\gamma_i|$ is constant and then the scattering invariance is manifest only for those polarizations on the same latitude circle of the Poincar\'{e} sphere (the special scenario of linear polarizations has already been studied in Refs.~\cite{YANG_2020_ACSPhotonics_Electromagnetic,YANG_2020_ArXiv200610629Phys._Scattering}); (ii) the optical activities are eliminated to ensure the scattering invariance for arbitrary polarizations. The elimination of optical activities means identical responses of the scattering systems for incident waves \textbf{L} and \textbf{R}: invariant cross sections of extinction, scattering or absorption require respectively $\rm {C}_{\rm{ext,sca,abs}}^L=\rm {C}_{\rm{ext,sca,abs}}^R$; invariant angular scattering patterns require $\rm{I}_{s}^{L}=\rm{I}_{s}^{R}$ along any scattering direction.  We emphasize that invariance of scattering cross sections are very different from the invariance of angular scattering patterns: the latter would always secure the former, while the opposite is wrong. This has been discussed already in Ref.~\cite{YANG_2020_ArXiv200613466Phys._Symmetry}, where the sole rotation symmetry produces invariant $\rm {C}_{\rm{sca}}$ for polarizations on the same latitude circle, but variant angular scattering patterns. Basically, the scattering invariance is manifest only when the scattering is integrated among all directions~\cite{YANG_2020_ArXiv200613466Phys._Symmetry}.

\begin{figure}[tp]
\centerline{\includegraphics[width=8.2cm]{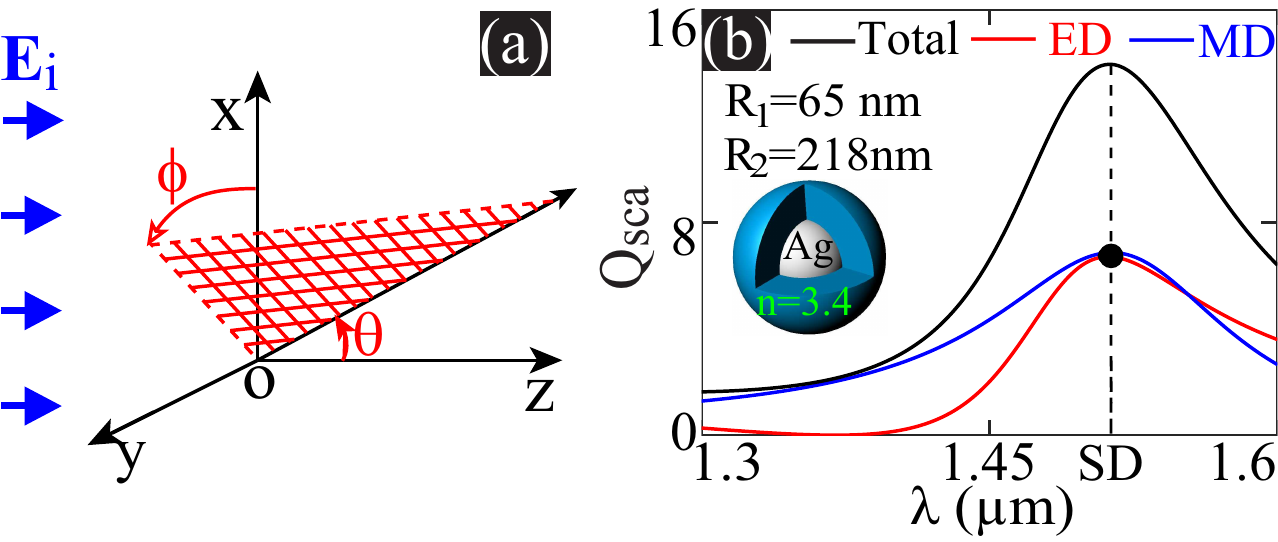}} \caption {\small (a) The scattering configuration is studied within the coordinate system parameterized by polar angles ($\theta$ and $\phi$) and Cartesian coordinates $x,y,z$. Unless specified otherwise, the incident plane wave is propagating along $z$ axis. (b) Scattering efficiency spectra (both total scattering and partial contributions from ED and MD) for the Ag core-dielectric ($n=3.4$) shell spherical particle ($R_1$ = 65 nm and $R_2$ = 218 nm).  The resonant spectral position (self-duality position) is marked at ${\lambda_\mathbf{SD}} = 1512$ nm.}\label{fig1}
\end{figure} 
\begin{figure}[tp]
\centerline{\includegraphics[width=8.7cm]{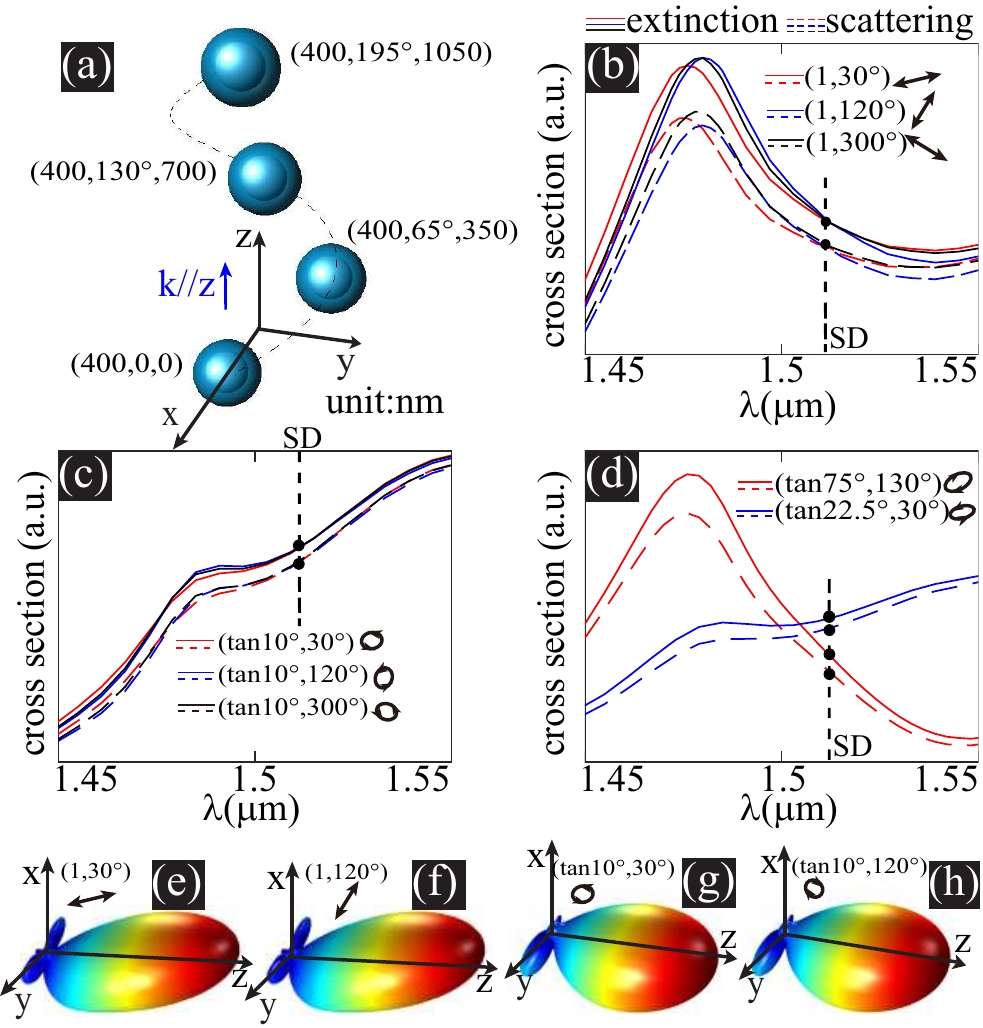}} \caption{\small  (a) A  scattering configuration consisting of four self-dual core-shell particles  that exhibits no other spatial symmetry. Location parameters specified in the figure are ($r=\sqrt{x^2+y^2},\phi,z$), and the unit for $r$ amd $z$ is nanometer. Extinction and scattering spectra for:  (b) linear polarizations of different orientations ($|\gamma_i|=1$);  (c) elliptical polarizations of different orientations while on the same lattitue circle of the Poincar\'{e} sphere [$|\gamma_i|=\tan(10^\circ$)]; two randomly chosen polarizations not on the same latitude circle in (d). The angular scattering patterns at the marked self-duality (SD) points in (b) and (c) are shown in (e)-(h). The incident polarizations are labelled by circular ratio parameters ($|\gamma_i|$, Arg$(\gamma_i)$) and the corresponiding polarization ellipses, as is also the case in Figs.~\ref{fig3} and ~\ref{fig4}.}
\label{fig2}
\end{figure}

\subsection{Invariant Scattering Properties Protected by Sole Duality Symmetry}

As a next step, to further verify what has been stated above in Section~\ref{section3-1}, we turn to specific demonstrations based on nonmagnetic self-dual core-shell particle clusters. We place our scattering configuration in the coordinate system shown in Fig. \ref{fig1}(a), and the employed fundamental building block is Ag core-dielectric shell (refractive index $n=3.4$) spherical particle, the self-duality of which at the resonant wavelength ${\lambda_\mathbf{SD}} = 1512$~nm is guaranteed by a pair of electric and magnetic dipoles (ED and MD) of equal strength supported~\cite{Wheeler2006_PRB,Liu2012_ACSNANO,YANG_2020_ACSPhotonics_Electromagnetic,YANG_2020_ArXiv200610629Phys._Scattering}. This is confirmed by the scattering efficiency (scattering cross section divided by the cross section of the particle) spectra (both total scattering and those contributed by individual ED and MD) shown in Fig. \ref{fig1}(b), where the self-duality resonant position is also marked.  As is the case throughout this work: the optical constants of silver are adopted from the experimental data listed in Ref.~\cite{Johnson1972_PRB}; the geometric parameters of scattering clusters are specified in the figure only; scattering properties are obtained through analytical calculations~\cite{Bohren1983_book} for the single particle [shown in Fig. \ref{fig1}(b)] and through numerical calculations (Comsol Multiphysics) for particle clusters.

\begin{figure}[tp]
\centerline{\includegraphics[width=8.5cm]{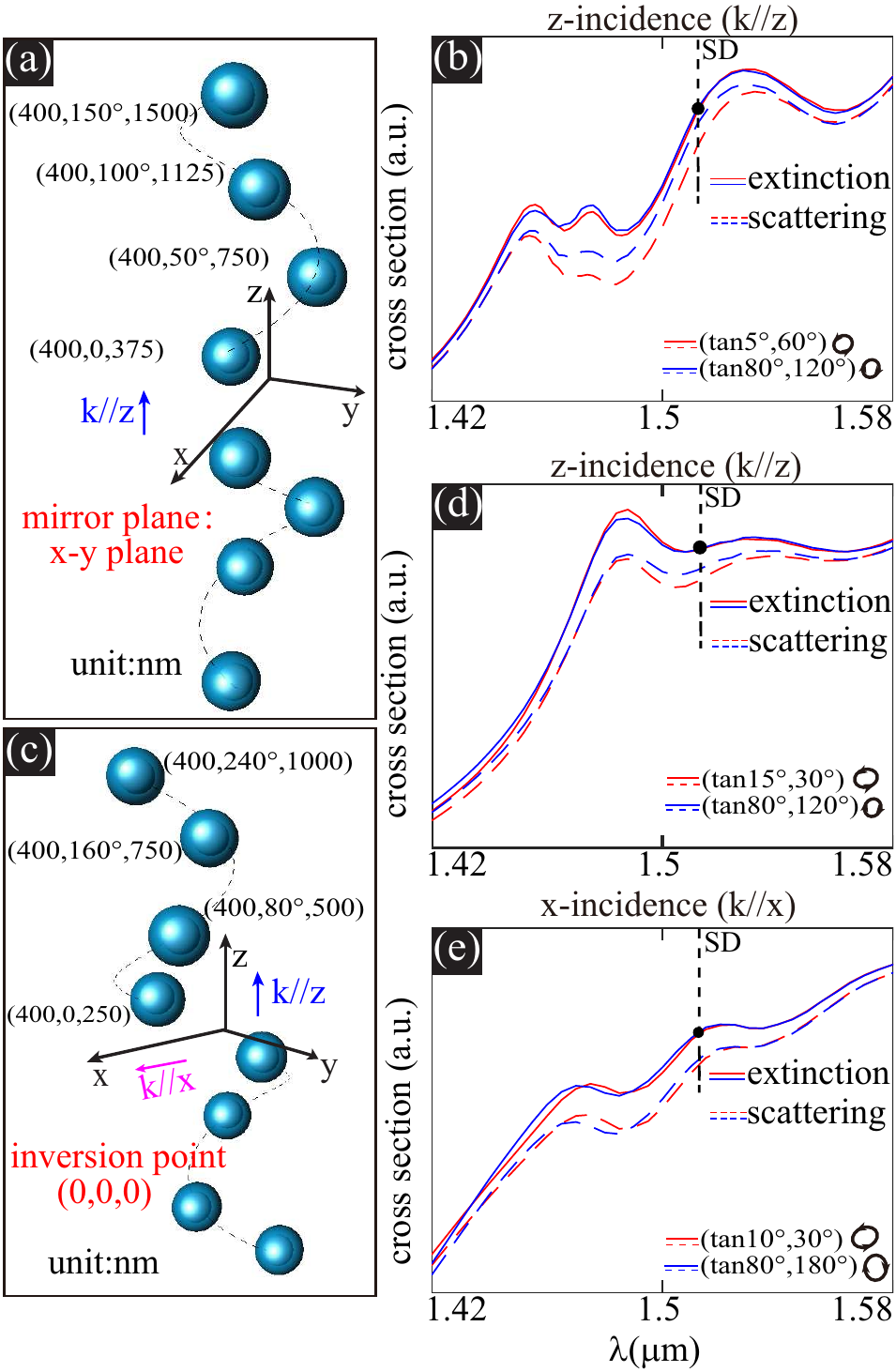}} \caption{\small Two scattering configurations consisting of eight self-dual core-shell particles each that exhibit extra perpendicular-mirror symmetry in (a) and inversion symmetry in (c). The two clusters are constructed by combing the cluster shown in Fig.~\ref{fig2}(a) with its perpendicular mirror-imaging (mirror position $z=0$) or point-imaging (inversion point position $x=y=z=0$) counterparts, respectively (location parameters specified in the figure).  The extinction and scattering spectra are shown respectively in (b) and (d), for two randomly chosen polarizations not on the same latitude circle of the Poincar\'{e} sphere. For the configuration in (c), another set of spectra for two arbitrarily chosen polarizations are shown in (e), for incident wave propagating along $\mathbf{x}$-axis.}\label{fig3}
\end{figure}

As has been mentioned already briefly in the introduction section, for a particle cluster consisting of such self-dual core-shell particles, the self-duality of the cluster is inherent, which is independent of the particle distributions or the wave incident directions. This is contrastingly different from spatial symmetries, where the symmetry of the whole scattering configuration is different from that of the scattering body, since the incident wave also has to be taken into consideration~\cite{CHEN_2020_Phys.Rev.Research_Scatteringa}. It means that the principles revealed in the Section~\ref{section3-1} can be directly applied to arbitrary self-dual clusters with arbitrary incident directions, based on which invariant scattering properties can be obtained for all polarization on the same latitude circle of the Poincar\'{e} sphere.

One such scattering configuration is shown in Fig.~\ref{fig2}(a), which exhibits no other spatial symmetries.  The cross section invariance at the self-duality wavelength ${\lambda_\mathbf{SD}}= 1512$~nm (only extinction and scattering cross section spectra are shown; absorption spectra can be obtained through a direct subtraction) are verified in:  Fig.~\ref{fig2}(b) for linear polarizations and Fig.~\ref{fig2}(c) for elliptical polarizations on the same latitude circle of the Poincar\'{e} sphere. For two polarizations not on the same latitude circle, the invariance would be broken, as is shown clearly in Fig.~\ref{fig2}(d).  The invariant scattering patterns at the self-duality wavelength [marked in Figs.~\ref{fig2}(b) and (c)] are shown in Figs.~\ref{fig2}(e) and (f) for linear polarizations, and in Figs.~\ref{fig2}(g) and (h) for elliptical polarizations. The invariance broken of scattering patterns for the two arbitrary polarizations [see Fig.~\ref{fig2}(d)] are not further demonstrated, as variant scattering cross sections must correspond to different scattering patterns. It is worth noting that the scattering invariance shown in Figs.~\ref{fig2}(b) and (c) are present not only at the designed self-duality frequency, but rather across a spectral regime, where the ED and MD supported overlap with each other of  almost the same strength.  Far from this regime where the self-duality of the consisting particle is lost (the ED and MD are of contrastingly different amplitudes), the invariance would be broken [see Figs.~\ref{fig2}(b) and (c)]. This feature is also manifest in Figs.~\ref{fig3}(b,d,e) and Fig.~\ref{fig4}(b), which  will be elaborated on later.

\subsection{Invariant Scattering Properties Protected by Joint Duality-Mirror (Perpendicular to Incident Direction) or Joint Duality-Inversion Symmetries}

The theoretical analysis outlined in Section~\ref{section3-1} elucidates that to extend the scattering invariance to cover the whole Poincar\'{e} sphere, we have to introduce extra symmetries to eliminate the corresponding optical activities. That is to say,  all-polarization independent extinction (scattering, or absorption) is achievable only when extinction (scattering, or absorption) activity is absent respectively, with the precondition that there is self-duality of the scattering system.

\begin{figure}[tp]
\centerline{\includegraphics[width=8.5cm]{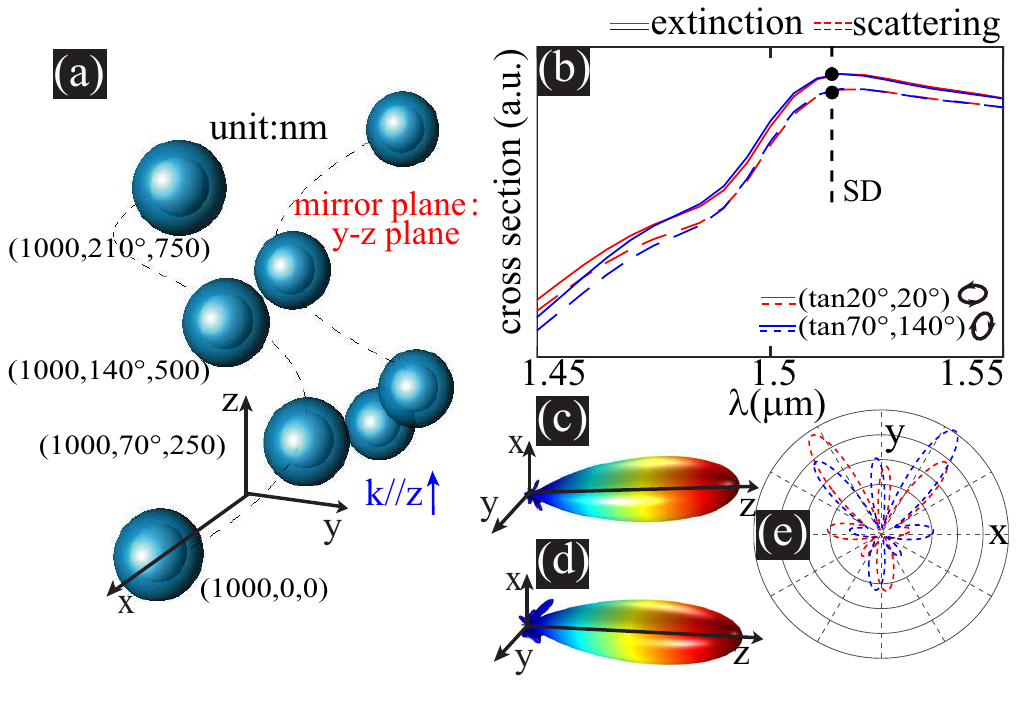}} \caption{\small (a) A scattering configuration consisting of eight self-dual core-shell particles that exhibit extra parallel-mirror  symmetry. The scattering cluster is  a combination of cluster shown in Fig.~\ref{fig2}(a) and its parallel mirror-imaging (mirror position $x=0$) counterpart (location parameters specified in the figure).  The extinction and scattering spectra are shown in (b), for two randomly chosen polarizations not on the same latitude circle of the Poincar\'{e} sphere. The  angular scattering patterns (including the two dimensional  patterns on the \textbf{x}-\textbf{y} planes)  at the self-duality marked point in (b) are shown in (c)-(e).}
\label{fig4}
\end{figure}

We have recently proved that the law of reciprocity and parity conservation can intrinsically eliminate the extinction activity (but not the scattering and absorption activities) when extra mirror (perpendicular to the incident direction) or inversion symmetry is introduced~\cite{CHEN_2020_Phys.Rev.Research_Scatteringa}. Two such scattering configurations are shown in Figs.~\ref{fig3}(a) and (c), which are basically the scattering configuration shown in Fig.~\ref{fig2}(a) combined with its perpendicular mirror-imaging or point-imaging counterparts, respectively.  Besides self-duality, the two scattering configurations exhibit respectively perpendicular mirror and inversion symmetry. The extinction invariance is confirmed in Figs.~\ref{fig4}(b) and (d), with two randomly chosen polarizations not on the same latitude circle for each scenario.  Compared to the mirror symmetry of the scattering configuration that imposes a stringent restriction for the incident direction (has to be perpendicular to the mirror in this case), the inversion symmetry is present for arbitrary incident directions, as is also the case for the extinction invariance. This incident direction independent invariance, for the configuration in  Fig.~\ref{fig3}(c) with extra inversion symmetry, is further verified in Fig.~\ref{fig3}(e) for another incident direction (\textbf{k} $\|$ \textbf{x}).  Since the extra perpendicular mirror or inversion symmetry cannot eliminate the scattering or absorption activities~\cite{CHEN_2020_Phys.Rev.Research_Scatteringa}, the absence of scattering and absorption invariance is manifest in all scenarios in  Figs.~\ref{fig3}(b), (d) and (e). As has been discussed already, the absence of scattering cross section invariance must be accompanied by different angular scattering patterns, and thus there is no need to further demonstrate the variance of angular scattering patterns.

We further note that in this section we have introduced either mirror or inversion symmetries only. A combination of them (such as those improper rotation symmetries~\cite{BARRON_2009__Molecular}) will also extinguish the extinction activity (but not the scattering or absorption activity)~\cite{CHEN_2020_Phys.Rev.Research_Scatteringa}, and thus could be employed to secure invariant extinction cross sections for arbitrary polarizations too.

\subsection{Invariant Scattering Properties Protected by Joint Duality-Mirror (Parallel to Incident Direction) Symmetries}
As has been clarified in Section~\ref{section3-1},  to obtain invariance of all cross sections for arbitrary polarizations requires the eliminations of all extinction, scattering and absorption activities.  This can be directly realized through another mirror (parallel to the incident direction) symmetry, where the law of parity conservation secures the same responses of the scattering bodies for incident LCP and RCP waves~\cite{BARRON_2009__Molecular,CHEN_2020_Phys.Rev.Research_Scatteringa}.   A self-dual scattering configuration with such parallel mirror symmetry is shown in Fig.~\ref{fig4}(a), which is basically the scattering configuration shown in Fig.~\ref{fig2}(a) combined with its parallel mirror-imaging counterpart.  The cross section invariance of extinction, scattering and absorption for arbitrary polarizations can be confirmed by Fig.~\ref{fig4}(b), where the scattering and extinction spectra for two randomly chosen polarizations are shown. 

As for the angular the scattering pattern, the parity conservation secures the parallel mirror symmetry of $\rm{I}_{s}^{L}$ and $\rm{I}_{s}^{R}$~\cite{BARRON_2009__Molecular,CHEN_2020_Phys.Rev.Research_Scatteringa,BARRON_2009__Molecular}: $\rm{I}_{s}^{L}(\phi,\theta)=\rm{I}_{s}^{R}(\pi-\phi,\theta)$ [if the mirror plane is \textbf{y}-\textbf{z} plane as shown in Fig.~\ref{fig4}(a)].  Since the invariance of angular scattering patterns required that $\rm{I}_{s}^{L}(\phi,\theta)=\rm{I}_{s}^{R}(\phi,\theta)$ for all scattering angles [refer to Eq.~(\ref{incident-arbitrary})], the joint parallel mirror-duality symmetries cannot secure the invariance of scattering patterns.  The variance of angular scattering patterns has been further verified in  Figs.~\ref{fig4}(c-e), where the patterns at the designed self-duality wavelength ${\lambda_\mathbf{SD}} = 1512$~nm [also marked in Fig.~\ref{fig4}(b)] are shown.  Here for better visibility of the contrast, two dimensional (2D) patterns on the \textbf{x}-\textbf{y} plane are also presented.

\section{Conclusions and Discussions}

To conclude, we reveal comprehensively how to exploit spatial and electromagnetic duality symmetries simultaneously to deliver invariant scattering properties (including cross sections of extinction, scattering, absorption) for arbitrary polarized incident waves. Sole self-duality of the scattering system ensures, regardless of geometric shapes or incident directions,  invariant scattering properties for all polarizations on the same latitude circle of the Poincar\'{e} sphere ($|\gamma_i|$ is constant) only.  It is further revealed that further introduction of extra mirror and/or inversion symmetries can extend such invariance to cover the whole Poincar\'{e} sphere, obtaining invariant scattering properties for arbitrary polarizations. The core results of this work shown in  Eqs.~(\ref{angular-intensity})-(\ref{absorption}) are valid not only for full polarized waves on the Poincar\'{e} sphere (as we have discussed in this work), but also for those partially polarized or even unpolarized waves within the Poincar\'{e} sphere.  Here for specific demonstrations, we have employed the elementary self-dual nonmagnetic particles supporting ED and MD only. The principles revealed are symmetry generic and thus applicable for any self-dual systems,  including but not limited to self-dual particles supporting equal higher-order multipoles~\cite{jahani_alldielectric_2016,KUZNETSOV_Science_optically_2016,LIU_2018_Opt.Express_Generalized} or even self-dual magnetic nonreciprocal particle clusters~\cite{FERNANDEZ-CORBATON_2013_Opt.ExpressOE_Forwarda,ZAMBRANA-PUYALTO2013Opt.Lett.,YANG_2020_ACSPhotonics_Electromagnetic}. The invariance we have obtained is inherent, which is protected by fundamental laws of reciprocity, and helicity and parity conservations, and thus would be robust against any symmetry-preserving perturbations.   The joint symmetry principles we have disclosed can provide novel insights for manipulating various sorts of symmetry-dictated light matter interactions, shedding new light on symmetry-related fundamental investigations (in fields such as topological and/or parity-time symmetric non-hermitian optics), and delivering extra opportunities for optical device designs.

\section*{acknowledgement}
We acknowledge the financial support from National Natural Science
Foundation of China (Grant No. 11874026, 11404403 and 11874426), and the Outstanding Young Researcher Scheme of National University of Defense Technology. W. L. thanks
Dr. Ivan Fernandez-Corbaton for helpful correspondences.



\begin{thebibliography}{10}
\newcommand{\enquote}[1]{``#1''}

\bibitem{FEYNMAN_2011__Feynmanb}
R.~P. Feynman, R.~B. Leighton, and M.~Sands, \emph{The {{Feynman Lectures}} on
  {{Physics}}, {{Vol}}. {{I}}: {{The New Millennium Edition}}: {{Mainly
  Mechanics}}, {{Radiation}}, and {{Heat}}} ({Basic Books}, {New York}, 2011),
  50th ed.

\bibitem{LANDAU_1984__Electrodynamicsb}
L.~D. Landau, L.~P. Pitaevskii, and E.~M. Lifshitz, \emph{Electrodynamics of
  {{Continuous Media}}: {{Volume}} 8} ({Butterworth-Heinemann}, {Amsterdam
  u.a}, 1984), 2nd ed.

\bibitem{JACKSON_1998__Classical}
J.~D. Jackson, \emph{Classical {{Electrodynamics Third Edition}}} ({Wiley},
  {New York}, 1998), 3rd ed.

\bibitem{Joannopoulos2008_book}
J.~D. Joannopoulos, \emph{Photonic {{Crystals}} : {{Molding}} the {{Flow}} of
  {{Light}}} (Princeton University, Princeton, 2008).

\bibitem{BARRON_2009__Molecular}
L.~D. Barron, \emph{Molecular {{Light Scattering}} and {{Optical Activity}}}
  ({Cambridge University Press}, 2009).

\bibitem{Bohren1983_book}
C.~F. Bohren and D.~R. Huffman, \emph{Absorption and Scattering of Light by
  Small Particles} (Wiley, 1983).

\bibitem{LIU_2005__Photonic}
J.-M. Liu, \emph{Photonic {{Devices}}} ({Cambridge University Press},
  {Cambridge ; New York}, 2005).

\bibitem{Hopkins2013_nanoscale}
B.~Hopkins, W.~Liu, A.~E. Miroshnichenko, and Y.~S. Kivshar, \enquote{Optically
  isotropic responses induced by discrete rotational symmetry of nanoparticle
  clusters,} Nanoscale \textbf{5}, 6395 (2013).

\bibitem{CHEN_2020_Phys.Rev.Research_Scatteringa}
W.~Chen, Q.~Yang, Y.~Chen, and W.~Liu, \enquote{Scattering activities bounded
  by reciprocity and parity conservation,} Phys. Rev. Research \textbf{2},
  013277 (2020).

\bibitem{YANG_2020_ArXiv200613466Phys._Symmetry}
Q.~Yang, W.~Chen, Y.~Chen, and W.~Liu, \enquote{Symmetry {{Protected Invariant
  Scattering Properties}} for {{Arbitrary Polarizations}},} arXiv:2006.13466
  (2020).

\bibitem{FERNANDEZ-CORBATON_2013_Phys.Rev.Lett._Electromagnetica}
I.~{Fernandez-Corbaton}, X.~{Zambrana-Puyalto}, N.~Tischler, X.~Vidal, M.~L.
  Juan, and G.~{Molina-Terriza}, \enquote{Electromagnetic {{Duality Symmetry}}
  and {{Helicity Conservation}} for the {{Macroscopic Maxwell}}'s
  {{Equations}},} Phys. Rev. Lett. \textbf{111}, 060401 (2013).

\bibitem{MOHAMMADIESTAKHRI_2020_Phys.Rev.Lett._Electromagnetic}
N.~Mohammadi~Estakhri, N.~Engheta, and R.~Kastner, \enquote{Electromagnetic
  {{Funnel}}: {{Reflectionless Transmission}} and {{Guiding}} of {{Waves}}
  through {{Subwavelength Apertures}},} Phys. Rev. Lett. \textbf{124}, 033901
  (2020).

\bibitem{YANG_2020_ACSPhotonics_Electromagnetic}
Q.~Yang, W.~Chen, Y.~Chen, and W.~Liu, \enquote{Electromagnetic duality
  protected scattering properties of nonmagnetic particles,} ACS Photonics,
  Doi: 10.1021/acsphotonics.0c00555  (2020).

\bibitem{YANG_2020_ArXiv200610629Phys._Scattering}
Q.~Yang, W.~Chen, Y.~Chen, and W.~Liu, \enquote{Scattering and absorption
  invariance of nonmagnetic particles under duality transformations,} arXiv:
  2006.10629  (2020).

\bibitem{FERNANDEZ-CORBATON_2013_Phys.Rev.B_Role}
I.~Fernandez-Corbaton and G.~Molina-Terriza, \enquote{Role of duality symmetry
  in transformation optics,} Phys. Rev. B \textbf{88}, 085111 (2013).

\bibitem{RAHIMZADEGAN_2018_Phys.Rev.Applied_CoreShella}
A.~Rahimzadegan, C.~Rockstuhl, and I.~{Fernandez-Corbaton},
  \enquote{Core-{{Shell Particles}} as {{Building Blocks}} for {{Systems}} with
  {{High Duality Symmetry}},} Phys. Rev. Applied \textbf{9}, 054051 (2018).

\bibitem{FERNANDEZ-CORBATON_2013_Opt.ExpressOE_Forwarda}
I.~{Fernandez-Corbaton}, \enquote{Forward and backward helicity scattering
  coefficients for systems with discrete rotational symmetry,} Opt. Express
  \textbf{21}, 29885--29893 (2013).

\bibitem{ZAMBRANA-PUYALTO2013Opt.Lett.}
X.~Zambrana-Puyalto, I.~Fernandez-Corbaton, M.~L. Juan, X.~Vidal, and
  G.~Molina-Terriza, \enquote{Duality symmetry and {{Kerker}} conditions,} Opt.
  Lett. \textbf{38}, 1857--1859 (2013).

\bibitem{CHEN_ACSOmega_Global}
W.~Chen, Q.~Yang, Y.~Chen, and W.~Liu, \enquote{Global {{Mie Scattering}}:
  {{Polarization Morphologies}} and the {{Underlying Topological Invariant}},}
  ACS Omega \textbf{5}, 14157--14163 (2020).

\bibitem{FEYNMAN_2011__Feynmanb3}
R.~P. Feynman, R.~B. Leighton, and M.~Sands, \emph{The {{Feynman Lectures}} on
  {{Physics}}, {{Vol}}. {{III}}: {{Quantum Mechanis}}} ({Basic Books}, {New
  York}, 2011), 50th ed.

\bibitem{BERRYM.V._2003_Proceeding_optical}
{Berry M. V.} and {Dennis M. R.}, \enquote{The optical singularities of
  birefringent dichroic chiral crystals,} Proc. R. Soc. Lond. A \textbf{459},
  1261--1292 (2003).

\bibitem{Wheeler2006_PRB}
M.~S. Wheeler, J.~S. Aitchison, and M.~Mojahedi, \enquote{Coated nonmagnetic
  spheres with a negative index of refraction at infrared frequencies,} Phys.
  Rev. B \textbf{73}, 045105 (2006).

\bibitem{Liu2012_ACSNANO}
W.~Liu, A.~E. Miroshnichenko, D.~N. Neshev, and Y.~S. Kivshar,
  \enquote{Broadband unidirectional scattering by magneto-electric core-shell
  nanoparticles,} ACS Nano \textbf{6}, 5489 (2012).

\bibitem{Johnson1972_PRB}
P.~B. Johnson and R.~W. Christy, \enquote{Optical constants of the noble
  metals,} Phys. Rev. B \textbf{6}, 4370 (1972).

\bibitem{jahani_alldielectric_2016}
S.~Jahani and Z.~Jacob, \enquote{All-dielectric metamaterials,} Nat.
  Nanotechnol. \textbf{11}, 23--26 (2016).

\bibitem{KUZNETSOV_Science_optically_2016}
A.~I. Kuznetsov, A.~E. Miroshnichenko, M.~L. Brongersma, Y.~S. Kivshar, and
  B.~Luk'yanchuk, \enquote{Optically resonant dielectric nanostructures,}
  Science \textbf{354}, aag2472 (2016).

\bibitem{LIU_2018_Opt.Express_Generalized}
W.~Liu and Y.~S. Kivshar, \enquote{Generalized {{Kerker}} effects in
  nanophotonics and meta-optics {{[Invited]}},} Opt. Express \textbf{26},
  13085--13105 (2018).

\end{thebibliography}

\end{document}